\documentclass[aps,notitlepage]{revtex4-1}

\usepackage[utf8]{inputenc}
\usepackage[english]{babel}

\usepackage{amsmath}
\usepackage{amsfonts}
\usepackage{amssymb}
\usepackage{graphicx}
\usepackage{epstopdf}
\usepackage{subfigure}
\usepackage{pxfonts}

\def\vec#1{\mathbf{#1}}

\def\vep{\varepsilon}
\def\uI{\overline{\overline{I}}}
\def\ubI{\overline{\overline{\mathbf{I}}}}
\def\u0{\overline{\overline{0}}}
\def\ub0{\overline{\overline{\mathbf{0}}}}
\def\uz{\overline{\overline{\zeta}}}
\def\uzo{\overline{\overline{\zeta}}_{\omega}}

\def\um{\overline{\overline{\mu}}}
\def\umo{\overline{\overline{\mu}}_{\omega}}
\def\uep{\overline{\overline{\vep}}}
\def\uepo{\overline{\overline{\vep}}_{\omega}}

\def\ubG{\overline{\overline{\textbf{G}}}}

\def\ubg{\overline{\overline{\textbf{g}}}}
\def\ubt{\overline{\overline{\textbf{t}}}}
\def\ubh{\overline{\overline{\textbf{h}}}}

\def\ubC{\overline{\overline{\textbf{C}}}}
\def\ubD{\overline{\overline{\textbf{D}}}}

\def\p{\partial}
\def\pr{\star}
\def\Do{D_{\Omega}}
\def\Dom{\Delta\omega}

\def\vf{\vec{f}}

\def\vD{\vec{D}}
\def\vB{\vec{B}}
\def\vE{\vec{E}}
\def\vH{\vec{H}}

\def\vX{\vec{\Xi}}
\def\vr{\vec{r}}
\def\vR{\vec{R}}

\def\vk{\vec{k}}
\def\vj{\vec{j}}

\def\hbO{\hat{\textbf{O}}}

\def\vbF{\vec{\textbf{F}}}
\def\vbJ{\vec{\textbf{J}}}

\def\zt{\zeta_{t}}
\def\zz{\zeta_{z}}
\def\ept{\vep_{t}}
\def\epz{\vep_{z}}
\def\mt{\mu_{t}}
\def\mz{\mu_{z}}

\def\uC{\overline{\overline{C}}}
\def\uD{\overline{\overline{D}}}
\def\Re{\mathop{\rm Re}}
\def\Im{\mathop{\rm Im}}

\def\hvx{\hat{\vec{x}}}
\def\hvy{\hat{\vec{y}}}
\def\hvz{\hat{\vec{z}}}
\def\hvu{\hat{\vec{u}}}
\def\ubN{\overline{\overline{\mathbf{N}}}}
\def\ubA{\overline{\overline{\mathbf{A}}}}
\def\ubT{\overline{\overline{\mathbf{T}}}}

\def\x{\times}
\def\xx{\mathchoice{\raisebox{1.5pt}{$_\times^\times$}}
             {\raisebox{1.5pt}{$\scriptstyle_\times^\times$}}
             {\raisebox{1.0pt}{$\scriptscriptstyle_\times^\times$}}
             {\raisebox{0.5pt}{$\scriptscriptstyle_\times^\times$}}}

\def\sgn{\operatorname{sgn}}

\def\DA2{\triangle^{2}_{A}}
\def\kt2{k_{t}^{2}}
\def\kz2{k_{z}^{2}}

\def\Dt{\kappa_{t}\kappa^{\pr}_{t}-\kappa_{t}\kappa^{-1}_{z}\kt2}

\def\Dt(p){\kappa_{t}\kappa_{t}^{\pr}-\kappa_{t}^{[\pr]}\kappa_{z}^{[\pr]-1}\kt2}

\begin{document}

\title{Envelope Dyadic Green's Function for Uniaxial Metamaterials}

\author{Stanislav I. Maslovski}
\email{stas@co.it.pt, stanislav.maslovski@gmail.com}

\affiliation{Instituto de Telecomunica\c{c}\~{o}es\\
  Departamento de Engenharia Electrot\'{e}cnica,
  Universidade de Coimbra, P\'olo II,
  3030-290 Coimbra, Portugal}

\author{Hodjat Mariji}
\email{hitcm@co.it.pt, hodjat.mariji@uc.pt}

\affiliation{Instituto de Telecomunica\c{c}\~{o}es\\
  Departamento de Engenharia Electrot\'{e}cnica,
  Universidade de Coimbra, P\'olo II,
  3030-290 Coimbra, Portugal}

\begin{abstract}

  Based on the dyadic Green's function (DGF) method, we present a
  formalism to study the propagation of electromagnetic fields with
  slowly varying amplitude (EMFSVA) in dispersive anisotropic media
  with two dyadic constitutive parameters, the dielectric permittivity
  and the magnetic permeability. We find the matrix elements of the
  envelope DGFs by applying the formalism for uniaxial anisotropic
  metamaterials. We present the relations for the velocity of the
  EMFSVA envelopes which agree with the known definition of the group
  velocity in dispersive media. We consider examples of propagation of
  the EMFSVA passing through active and passive media with the Lorentz
  and the Drude type dispersions, demonstrating beam focusing in
  hyperbolic media and superluminal propagation in media with inverted
  population. The results of this paper are applicable to the
  propagation of modulated electromagnetic fields and slowly varying
  amplitude fluctuations of such fields through dispersive and
  dissipative (or active) anisotropic metamaterials. The developed
  approach can be also used for the analysis of metamaterial-based
  waveguides, filters, and delay lines.
	 
\end{abstract}

\maketitle

\section{Introduction}

Nowadays, there has been a growing interest in metamaterials (MMs)
which are artificial composites used in various branches of science
and engineering. After the pioneering works on the backward wave
propagation in media \cite{1}--\cite{4}, many scientists have focused
on the appealing features of MMs such as the negative refractive index
of active and passive MMs \cite{5,5a,6}, the diffraction-unlimited
imaging \cite{11}, and the remarkable control over the electromagnetic
fields \cite{8}. These features allowed for fabrication of flat lenses
\cite{19}, invisible cloaking devices \cite{12}, and perfect
electromagnetic absorbers \cite{9, 10}. MMs have been employed for
the modeling of general relativity effects with artificial black holes
\cite{7} and to achieve frequency-agile or multi-band operation
\cite{13}--\cite{15}, to mediate the repulsive Casimir force \cite{16},
and to tune microwave propagation with light \cite{18}. Another area
in which MMs may have a broad impact that has recently attracted
attention of science and technology is the near-field super-Planckian
radiative heat transfer \cite{20}--\cite{22}, with applications for
thermophotovoltaics \cite{23, 24}.

In some situations, the unusual dispersive properties of the MMs
result in superluminal or subluminal group velocities,
$V_g = \partial \omega/\partial k$, in which case this quantity can be
higher~\cite{Wang2000} or extremely lower than the speed of light in a
vacuum~\cite{26}. It can even become zero or negative
\cite{47}--\cite{28}. These effects in MMs, as well as the
negative refractive index, are well-accommodated within the framework
of the causality principle \cite{29}--\cite{32}.  In recent decades,
superluminal and subluminal group velocities have attracted attention
in nonlinear optics \cite{33}, quantum communication \cite{34}, photon
controlling and storage \cite{35}--\cite{37}, precision sensing
\cite{38}, high-speed optical switching \cite{39, 40}, broadband
electromagnetic devices and delay compensation circuits in
ultra-high-speed communication systems \cite{41}, and in high
resolution spectrometers \cite{42}.

In this article, based on the dyadic Green's function (DGF) method, we
present a self-consistent formalism to solve the Maxwell equations
written for the electromagnetic fields with slowly varying amplitude
(EMFSVA), when such fields propagate through a dispersive anisotropic
medium with known dyadics of permittivity ($\uep$) and permeability
($\um$). Generalization to the case of bianisotropic media is as well
possible. Here we focus on uniaxial anisotropic MMs, because they
allow for a closed-form analysis. This formalism can be used, e.g.\ to
study the propagation of electromagnetic fluctuations in
super-Planckian radiative heat transfer systems~\cite{22}, which is an
actively developing topic in the context of MM applications. The
uniaxial MMs are known for advantageous optical properties for sensing
\cite{43}, nonlinear optical applications \cite{44}, and spontaneous
emission control \cite{45}. In this work, we examine our formalism on
hyperbolic MMs and media with gain, in order to demonstrate the
applicability of the method to the wave processes in such media,
including the exotic superluminal processes in active media.

There is a significant body of literature on DGFs in anisotropic and
bianisotropic
media~\cite{Lee83,Lakhtakia89,Weiglhofer90,Weiglhofer93,Weiglhofer94,Lindell99,Olyslager01,Olyslager02}.
While in the most general case of bianisotropic medium there is no
closed-form representation for the DGF, in the case of uniaxial
magnetodielectrics, the time-harmonic DGF can be expressed through a
pair of scalar electric and magnetic Green's functions. When
interested in the EMFSVA in such media, the standard approach would be
to start from such closed-form representations and expand the
frequency-dependent parameters (such as propagation factors, etc.) in
these relations into series around the carrier frequency $\omega_0$,
with $\Omega = \omega - \omega_0$ being the small parameter. Although
this would constitute a sound approach for the propagation of the
EMFSVA in uniaxial media, in this article we develop a different
method, which can be later extended to general bianisotropic media.

We start from introducing the slowly varying amplitudes (SVAs) of
electromagnetic fields and write the Maxwell equations in terms of
these quantities. With these equations, the dynamics of the EMFSVA can
be studied directly, with material dispersion-related quantities like
${\p(\omega\uep)\over\p\omega}$ and ${\p(\omega\um)\over\p\omega}$
appearing in these equations in a natural way. The envelope dyadic
Green's function (EDGF) is then introduced as the solution of the
corresponding dyadic equation with a point-like source. Note that, in
contrast to the conventional DGF for the time-harmonic sources, the
EDGF is a wavelet-type dyadic function that describes propagation of
the amplitude fluctuations of quasi-monochromatic signals. These
fluctuations propagate with a certain velocity that has the meaning of
the group velocity. Despite being introduced in our formalism through
a different way, this velocity agrees with the classical definition of
the group velocity in dispersive media. Although not considered in
this article, the presented formulation allows also for a direct
finite-difference time-domain-based numerical solution of the dynamic
equations for the EMFSVA, which can be useful, e.g.~when studying
propagation of wave packets in layered MMs.

In order to test our formalism, we consider the propagation of the
EMFSVA through a hyperbolic medium and through an active medium,
e.g.~the inverted population $^{132}$Xe gas, which are sandwiched
between two passive dielectric layers. In the former case, the
paraxial propagation of the EMFSVA leads to the well-known negative
refraction effect~\cite{Lindell01,Smith04}, while in the latter case
one can observe the peculiar phenomenon of negative group velocity
associated with superluminal propagation~\cite{47}. It is worth to
mention that the paraxial propagation of the EMFSVA can be used to
study the near-field thermal radiation effects in extremely
anisotropic media, e.g.~in arrays of aligned carbon
nanotubes~\cite{Liu2013}.

This article is organized as follows: In Sec.~\ref{sec:tools}, we
briefly set up the main tools to be used in our formalism for the
EMFSVA; in Sec.~\ref{sec:edgf}, which employs the EDGF technique, the
main formalism is presented; in Sec.~\ref{sec:matrix}, we obtain the
EDGF matrix elements for the uniaxial anisotropic MMs; in
Secs.~\ref{paraxial} and \ref{layered}, we develop the paraxial
approximation for the EDFG; in Sec.~\ref{sec:exampl}, we study the
above-mentioned effects associated with the propagation of the EMFSVA
in active and passive media. Finally, we draw conclusions in
Sec.~\ref{sec:concl}.
  
\section{\label{sec:tools}Main EMFSVA Tools and Definitions}

In order to investigate the propagation of the EMFSVA in a dispersive
medium, we consider the Maxwell equations for the time-dependent
electromagnetic fields and sources written as follows
\begin{align}\label{1}
    \nabla \cdot \vD=\rho^{E}, \qquad \nabla \cdot \vB=\rho^{M},\qquad
    \nabla\times\vE=-\p_{t}\vB-\vj^{M}, \qquad \nabla\times\vH=\p_{t}\vD+\vj^{E},
\end{align}
where the vectors $\vE$, $\vH$, $\vD$, and $\vB$ are the electric
field, the magnetic field, the electric displacement, and the magnetic
flux density, respectively, and $\p_{t} = {\p \over \p t}$. In
Eq.~(\ref{1}), $\rho^{E(M)}$ and $\vj^{E(M)}$, the electric (magnetic)
charge and current densities, respectively, are related by the
continuity equation, $\nabla\cdot\vj^{E(M)} + \p_{t} \rho^{E(M)}=0$.
The electromagnetic constitutive equation is given by
\begin{equation}\label{2}
\vX=\uz\cdot\vf,
\end{equation}
where the dyadic integro-differential operator $\uz = \uep$ ($\um$)
represents the dispersive anisotropic permittivity
(permeability) of the material, which relates $\vX= \vD\,(\vB)$ to
$\vf= \vE\,(\vH)$.

The time-dependent electromagnetic field can be expanded into the monochromatic spectral components as follows
\begin{equation}\label{3}
\vf(t)={1 \over 2\pi}\int_{-\infty}^{+\infty} \vf_\omega(\omega)e^{-i\omega t}d\omega.
\end{equation}
On the other hand, considering the time-harmonic
fields with SVAs, it is acceptable that the most
spectral energy is concentrated in a narrow band $\Delta\omega$ around
$\omega_{0}$, the carrier frequency, with
$\Delta\omega \ll \omega_ {0}$. Thus, we can define the EMFSVA, $\vf_m(t)$, as follows 
\begin{align}
  \vf(t)={1 \over 2}\vf_{m}(t)e^{-i\omega_{0}t} + \mbox{c.c.},\label{eq:ft} \qquad
  \vf_{m}(t)={1 \over \pi} \int_{-\Delta\omega/2}^{+\Delta\omega/2}\vf_\omega(\omega_{0}+\Omega)e^{-i\Omega t}d\Omega,
\end{align} 
where c.c.\ is the abbreviation for the complex conjugate of the
previous term and $\Omega=\omega-\omega_{0}$. We also assume that
$\uz_\omega$, the Fourier transform of $\uz$, has appreciably smooth
behavior in the narrow band $\Delta\omega$, which is a reasonable
assumption for applications of our interest. So, expanding
$\uz_\omega$ around $\omega_{0}$ and keeping only the two first terms,
we obtain
\begin{equation}\label{5}
\uz_\omega=\uz_\omega\big|_{\omega_0} + (\p_{\omega}\uz_\omega)\big|_{\omega_{0}} \Omega,
\end{equation}  
where $\p_{\omega} = {\p \over \p\omega}$ and $|_{\omega_{0}}$ denotes
that the related quantity is computed at $\omega_{0}$. Using
Eqs.~(\ref{eq:ft})--(\ref{5}), we obtain $\p_{t}\vX$ as follows
\begin{align}\label{6}
\p_{t}\vX = e^{-i\omega_{0}t}(\p_{t} - i\omega_{0})\vX_{m}(t),\qquad 
\vX_{m}(t)=\left(\uz_\omega\big|_{\omega_0} + i(\p_{\omega}
            \uz_\omega)\big|_{\omega_0}\p_{t}\right)\cdot\vf_{m}(t) \equiv \uz_m\cdot\vf_{m}(t),
\end{align}
where the dyadic differential operator $\uz_m$ is
$\uz_m = \uz_\omega\big|_{\omega_0} + i(\p_{\omega}
\uz_\omega)\big|_{\omega_0}\p_{t}$, and we have to ignore the second
and higher orders of $\p_t$ in expressions involving this
operator. With having these tools at hand, in the next section we
proceed to the Green's function technique to solve the Maxwell
equations written in terms of the EMFSVA.
  
\section{\label{sec:edgf}Envelope DGF Technique}

Starting from Eq. (1) 
for the time-dependent electromagnetic fields and sources and using
Eqs.~(\ref{2})--(\ref{6}), the Maxwell 6-vector
equations for the EMFSVA assume the following operator form:
\begin{equation}\label{7}
\hbO\cdot\vbF_{m}=-i\vbJ_{m}.
\end{equation}
In the above equation, $\vbF_{m}$, the SVA of the 6-vector field,
$\vbJ_{m}$, the 6-vector current density, and $\hbO$, the
electromagnetic operator, are given by
\begin{equation}\label{8}
\vbF_{m}\dot{=}
\begin{bmatrix}
\vE_{m} \\
\vH_{m}
\end{bmatrix},\qquad
\vbJ_{m}\dot{=}
\begin{bmatrix}
\vj_{m}^{E} \\
\vj_{m}^{M}
\end{bmatrix},\qquad
\hbO\dot{=}
i\begin{bmatrix}
\left(\p_{t}-i\omega_{0}\right)\uep_m & -\nabla\times\uI \\
\nabla\times\uI & \left(\p_{t}-i\omega_{0}\right)\um_m 
\end{bmatrix},
\end{equation}
where $\uI$ is the identity dyadic. In Eq.~(\ref{8}), the differential operators $(\p_{t}-i\omega_{0})\uz_m$ are expanded as
\begin{equation}
  (\p_{t}-i\omega_{0})\uz_m = -i\omega_0\uz_\omega\big|_{\omega_0} +\p_\omega(\omega\uz_\omega)\big|_{\omega_0}\p_t,
\end{equation}
with the second order derivative term ignored, in line with the
assumption of the SVA.  We may introduce the envelope 6-dyadic Green's
function (EDGF), $\ubG(\vec{r},t;\vec{r}',t')$, for Eq.~(\ref{7}) such
that
\begin{equation}\label{9}
\hbO\cdot\ubG=-i
\delta(\vr-\vr')\tilde{\delta}(t-t')\ubI,
\end{equation}
where $\ubI$ is the identity 6-dyadic, and $\tilde{\delta}(\cdot)$,
similar in role to the Dirac delta $\delta(\cdot)$, emphasizes that we
consider the slowly varying sources and fields. Respectively,
$\tilde{\delta}(t)$ has a finite duration and
$|\tilde{\delta}(t)| < \infty$.  Eq.~(\ref{9}) means that the EDGF is
obtained by inverting the operator $\hbO$.

It is more convenient to work in the Fourier domain when the material
parameters are uniform in both space and time. In this case,
$\ubG(\vr,t;\vr',t') \equiv \ubG(\vr-\vr',t-t')$, and we can define
$\ubg(\vk,\Omega)$, the Fourier transform of the EDGF, by the
following relation
\begin{equation} \label{12}
\ubG(\vR, \tau) = {1 \over (2\pi)^{4}} \int
d\vk\int_{-\Delta\omega/2}^{+\Delta\omega/2}d\Omega\,\,\ubg(\vk,\Omega)
e^{i(\vk\cdot\vR-\Omega \tau)},
\end{equation}
where $\vR = \vr-\vr'$ and $\tau = t-t'$.
Recalling Eq. (\ref{9}) and using that, in the Fourier space,
$\p_{t}=-i\Omega$ and $\nabla=i\vk$, where $\vk$ is the wave vector,
we find that the 6-dyadic equation for $\ubg$ satisfies
\begin{equation}\label{13}
\hbO\cdot\ubg=-i\ubI, \qquad
\hbO\dot{=}
\begin{bmatrix}
\Do\uepo \qquad \vk\times\uI \\
-\vk\times\uI \qquad \Do\umo 
\end{bmatrix},
\end{equation}
where we have used operator-like notation for $D_\Omega$ such that
\begin{equation} \label{eq:Do}
  \Do\uzo\equiv\omega_{0}\uzo\big|_{\omega_0}+\Omega\,\p_{\omega}(\omega\uzo)\big|_{\omega_0},
\end{equation}
and the fact that
$\tilde{\delta}(\tau) =
(2\pi)^{-1}\int_{-\Delta\omega/2}^{+\Delta\omega/2}e^{-i\Omega\tau}\,d\Omega$.

Knowing the dispersive constitutive parameters of the medium and
inverting the operator $\hbO$ in Eq.~(\ref{13}), we can obtain $\ubg$
and, in turn, the EDGF. Afterward, for any given forms of the 6-vector
source functions $\vbJ_m$ with the frequency spectra fitting the
interval $\Omega \in [-{\Dom\over 2};+{\Dom\over 2}]$, the components
of the 6-vector $\vbF_{m}$ are obtained with the help of Kotelnikov's
theorem~\cite{Kotelnikov33} (see Appendix A) as follows
\begin{equation}\label{15}
\vbF_{m}\left(\vr,t\right)={2\pi\over\Dom}\sum_{n=-\infty}^{+\infty}\int
d\vr^{\prime}\,\ubG(\vr-\vr', t-t_n)\cdot
\vbJ_{m}(\vr^{\,\prime},t_n),
\end{equation}
where $t_n = {2\pi n\over\Dom}$. In the next section, we shall find
the matrix elements of the EDGF for uniaxial anisotropic media.
 
\section{\label{sec:matrix}EDGF Matrix Elements}

Here, we apply the EDGF technique for reciprocal uniaxial media to
obtain the matrix elements. For such media we can write
\begin{equation}\label{16}
\uz_\omega=\zt\uI_{t}+\zz\uI_{z}	
\end{equation}
with $\uI_{t}=\hvx\hvx+\hvy\hvy$ and $\uI_{z}=\hvz\hvz$ being the
projection dyadics in the transverse and axial (the main axis)
directions, respectively, along which the electromagnetic responses of
the medium are different. In order to obtain the EDGF of the uniaxial
medium, we impose this property of constitutive parameters to
decompose the operator $\hbO$ in Eq.~(\ref{13}) as
\begin{equation}\label{17}
\hbO\dot{=}\left[
\begin{array}{cc|cc}
  \Do\ept\uI_{t} & \u0 & \vk_{z}\times\uI_{t} & \vk_{t}\times\uI_{z}\\
  \u0 & \Do\epz\uI_{z} & \vk_{t}\times\uI_{t} & \u0 \\
  \hline
  -\vk_{z}\times\uI_{t} & -\vk_{t}\times\uI_{z}& \Do\mt\uI_{t}  & \u0 \\
  -\vk_{t}\times\uI_{t} & \u0 & \u0 & \Do\mz\uI_{z}
\end{array}
\right],
\end{equation}
where $\u0$ is the null dyadic. When writing the matrix elements in
Eq.~(\ref{17}), we keep in mind that
$\vk\times\uI=(\vk_{t}+\vk_{z})\times(\uI_{t}+\uI_{z})$, where
$\vk_{t} = \uI_{t}\cdot\vk$ and $\vk_{z} = \uI_{z}\cdot\vk$. Next, it is more
convenient if we replace the second row with the third one and also
the second column with the third one in the matrix representation of
$\hbO$ in Eq.~(\ref{17}) so that the tangential and axial components
of the constitutive parameters of the medium take place in separate
blocks.  When doing this, it is also necessary to respectively
rearrange the dyadic components of $\ubg$ and $\ubI$ in
Eq.~(\ref{13}).  After doing this and taking into account that
$\vk_t\x\uI_t = \uI_z\x\vk_t$, the EDGF is expressed through
the inverse of the reordered matrix of $\hbO$ as follows
\begin{align}\label{18}
\ubg = \left[
\begin{array}{c|c}
  \ubg^{tt} & \ubg^{tz}\\
  \hline
  \ubg^{zt} & \ubg^{zz}
\end{array}
\right] 
= - i\left[
\begin{array}{cc|cc}
  \Do\ept\uI_{t}        & \vk_{z}\times\uI_{t} &\u0                 & \vk_{t}\times\uI_{z}\\
  -\vk_{z}\times\uI_{t} & \Do\mt\uI_{t}        &-\vk_{t}\times\uI_{z}& \u0 \\
  \hline
  \u0                  & \uI_z\times\vk_{t}   & \Do\epz\uI_{z}     & \u0 \\
  -\uI_z\times\vk_{t}   & \u0                  & \u0                & \Do\mz\uI_{z}
\end{array}
\right]^{-1}\!\!\!\!\!.
\end{align}

Representation~(\ref{18}) allows for a straightforward splitting of
the fields into a pair of orthogonal polarizations: The
transverse-electric (TE or $s$-) polarization with vanishing axial component
of the electric field, and the transverse-magnetic (TM or $p$-) polarization
with vanishing axial component of the magnetic field. In order to
perform such splitting, $\uI_t$ in Eq.~(\ref{18}) is expanded as
$\uI_t = {\vk_t\vk_t\over k_t^2} + \uI_z\xx{\vk_t\vk_t\over k_t^2}$,
where $\xx$ denotes the dyadic double cross product:
$\vec{a}\vec{b}\xx\vec{c}\vec{d} =
\vec{a}\times\vec{c}\,\vec{b}\times\vec{d}$ and $k_t = |\vk_t|$. Then, after a rather
tedious but straightforward dyadic algebra, the matrix in
Eq.~(\ref{18}) can be inverted and the following result obtained:

\begin{align}\label{gtt}
  \ubg^{tt} =
  \begin{bmatrix}
    g^{tt,p}_{ee}{\vk_t\vk_t\over k_t^2} &
    g^{tt,p}_{em}{\vk_t\hvz\x\vk_t\over k_t^2}\\
    g^{tt,p}_{em}{\hvz\x\vk_t\vk_t\over k_t^2} &
    g^{tt,p}_{mm}{\hvz\hvz\xx\vk_t\vk_t\over k_t^2}
  \end{bmatrix} 
  \,+
  \begin{bmatrix}
    g^{tt,s}_{ee}{\hvz\hvz\xx\vk_t\vk_t\over k_t^2} &
    g^{tt,s}_{em}{\hvz\x\vk_t\vk_t\over k_t^2}\\
    g^{tt,s}_{em}{\vk_t\hvz\x\vk_t\over k_t^2} &
    g^{tt,s}_{mm}{\vk_t\vk_t\over k_t^2}
  \end{bmatrix},
\end{align}
\begin{equation}
  \ubg^{tz} = \left(\ubg^{zt}\right)^T = 
  \begin{bmatrix}
    g^{tz,p}_{ee}{\vk_t\hvz\over k_t} &
    g^{tz,s}_{em}{\hvz\x\vk_t\hvz\over k_t}\\
    g^{tz,p}_{me}{\hvz\x\vk_t\hvz\over k_t} &
    g^{tz,s}_{mm}{\vk_t\hvz\over k_t}
  \end{bmatrix},
\end{equation}
\begin{equation}\label{gzz}
  \ubg^{zz} =
  \begin{bmatrix}
    g^{zz,p}_{ee}\uI_z & \u0 \\
    \u0 & g^{zz,s}_{mm}\uI_z
  \end{bmatrix},
\end{equation}
where the 12 independent non-vanishing components of $\ubg$ are
expressed as follows (here and thereafter we use shorthand notation
$\alpha[\beta]$ for respective selection of either $\alpha$ or
$\beta$, where $\alpha$ and $\beta$ are isolated
terms or groups of indices):
\begin{equation} \label{22}
  \begin{gathered}
    g^{tt,p[s]}_{ee[mm]} = -i{d^{m[e]}_t-k_t^2/d^{e[m]}_z\over \kappa_{p[s]}^2-k_z^2},\qquad
    g^{tt,s[p]}_{ee[mm]} = -{id^{m[e]}_t\over \kappa_{s[p]}^2-k_z^2},\qquad
    g^{tt,s[p]}_{em} = [-]{ik_z\over \kappa_{s[p]}^2-k_z^2},
  \end{gathered}
\end{equation}
\begin{equation} \label{23}
  \begin{gathered}
    g^{tz,p[s]}_{ee[mm]} = {ik_tk_z/d^{e[m]}_z\over
      \kappa_{p[s]}^2-k_z^2}, \qquad g^{tz,p[s]}_{me[em]} = [-]{ik_td^{e[m]}_t/d^{e[m]}_z\over \kappa_{p[s]}^2-k_z^2}, \qquad
    g^{zz,p[s]}_{ee[mm]} = -i{(d^e_td^m_t-k_z^2)/d^{e[m]}_z\over
      \kappa_{p[s]}^2-k_z^2},
  \end{gathered}
\end{equation}
where, from Eq.~(\ref{eq:Do}),
\begin{equation}
  \begin{gathered}
    d^{e[m]}_l = a^{e[m]}_l+b^{e[m]}_l\Omega, \qquad
    a^{e[m]}_l =\omega_0\vep_l[\mu_l]\big|_{\omega_{0}},\qquad
    b^{e[m]}_l =\p_\omega(\omega\vep_l[\mu_l])\big|_{\omega_{0}},
  \end{gathered}\label{20}
\end{equation}
where the index $l$ is either $t$ or $z$, and
\begin{equation}
\kappa_{p[s]}=\sqrt{d^e_td^m_t-\left(d^{e[m]}_t/d^{e[m]}_z\right)k_t^2},
\quad \Im(\kappa_{p[s]}) \ge 0.\label{24}
\end{equation}

As can be seen from Eqs.~(\ref{22}) and~(\ref{23}), there are two
poles in $k_z$ for each polarization: $\pm \kappa_p$ for the TM case
and, similarly, $\pm \kappa_s$ for the TE one. Physically, the poles
$k_z = +\kappa_{p[s]}$ correspond to the waves propagating in the
halfspace $z - z' > 0$, and the poles with $k_z = -\kappa_{p[s]}$
correspond to the waves propagating at $z - z' < 0$.

Based on Eqs.~(\ref{22}) and~(\ref{23}), we can sort out the
components of $\ubg$ into two groups which correspond to the waves of
$p$- and $s$-polarization
\begin{equation}
  \ubg = {\ubN_p\over \kappa_p^2-k_z^2} + {\ubN_s\over \kappa_s^2-k_z^2},
\end{equation}
where $\ubN_{p[s]}$ are formed by the corresponding terms in the
numerators of Eqs.~(\ref{22}) and~(\ref{23}).
Respectively, when taking the inverse Fourier transform of $\ubg$ as
given by Eq.~(\ref{12}) and considering the integral over $dk_z$, we get
two categories of residues
\begin{equation}
  \begin{aligned}
    {1\over 2\pi}\int\limits_{-\infty}^{\infty}dk_z\,\ubg\,e^{i k_z Z}
    = -\sum_{\gamma = p,s}{i\over
      2\kappa_\gamma}\ubN_\gamma\big|_{k_z=\sgn(Z)
      \kappa_\gamma}e^{i\kappa_{\gamma}|Z|} 
    = \sum_{\gamma = p,s}\ubA_{\gamma,\sgn(Z)}
    e^{i\kappa_{\gamma}|Z|},
  \end{aligned}
\end{equation}
where $Z = z - z'$, and $\sgn(Z) = \pm 1$ is the sign of $Z$.

Next, before taking the integral over $d\Omega$, we
recall the approximation of the SVA, and expand
$\ubA_{\gamma,\pm 1}$ and $\kappa_{\gamma}$ around the point $\Omega=0$ and ignore
$O(\Omega^2)$ terms so that
\begin{equation}\label{defCandD}
\ubA_{\gamma,\pm 1}=\ubA_{\gamma,\pm 1}\big|_{\Omega=0} + \Omega
\left.{\p \ubA_{\gamma,\pm 1} \over \p \Omega} \right|_{\Omega=0}
\equiv \ubC_{\gamma,\pm 1} + \Omega\,\ubD_{\gamma,\pm 1},
\end{equation}
\begin{equation}\label{26}
\kappa_{\gamma}=\kappa_{\gamma}\big|_{\Omega=0} + \Omega \left.{\p
    \kappa_{\gamma} \over \p \Omega} \right|_{\Omega=0} \equiv
\kappa^{\gamma}_0 + \Omega/V^{\gamma}_{g},
\end{equation}
where the expressions for the components of
$\ubC_{\gamma,\pm 1}=\ubA_{\gamma,\pm 1}\big|_{\Omega=0}$ and
$\ubD_{\gamma,\pm 1}=\p \ubA_{\gamma,\pm 1}/\p
\Omega\,\big|_{\Omega=0}$ are given in Appendix B, and for the
propagation factor $\kappa^{\gamma}_0=\kappa_{\gamma}|_{\Omega=0}$ and
the complex group velocity
$V^{\gamma}_{g} =
{\left(\p\kappa_\gamma/\p\Omega\right)^{-1}_{\Omega=0}}$ we obtain
\begin{align}
\label{28}
\kappa^{p[s]}_{0} =& \sqrt{a^e_{t}a^m_{t}-{a^{e[m]}_{t} \over
  a^{e[m]}_{z}} \kt2},\\
V^{p[s]}_{g} =& { 2\kappa^{p[s]}_{0} \over
  (a_{t}^{e}b^{m}_{t}+a^{m}_{t}b^{e}_{t})
 - (b^{e[m]}_{t}a^{e[m]}_{z}-a^{e[m]}_{t}b^{e[m]}_{z}) \kt2 / {a^{e[m]}_{z}}^2 },\label{29}
\end{align}
for the $p$- [$s$-] polarization. It should be noted that ignoring the
second order terms in Eqs.~(\ref{defCandD}) and (\ref{26}) is
reasonable for the frequency intervals of our interest, where the
group velocity dispersion effects are relatively weak. In Appendix C,
we discuss this with more detail and also present a closer look at the
group velocity for a propagating envelope in an active and dispersive
medium.

With these approximations at hand, when performing the integration over $\Omega$ we obtain
\begin{multline}
  {1\over 2\pi}\!\!
  \int\limits_{-\Dom/2}^{+\Dom/2}\!\!\!d\Omega
  \sum_{\gamma = p,s}
  \left(
    \ubC_{\gamma,\sgn(Z)}+\Omega\,\ubD_{\gamma,\sgn(Z)}
  \right)
  e^{i\kappa^\gamma_0|Z|-i\Omega\tau_g^\gamma}
  ={\Dom\over 2\pi} \sum_{\gamma = p,s}
  e^{i\kappa^\gamma_0|Z|}\bigg[
    \ubC_{\gamma,\sgn(Z)}\,j_0
    \left({\Dom\tau_g^\gamma/2}\right) - {i\Dom\over 2}
    \ubD_{\gamma,\sgn(Z)}\,j_1\left({\Dom\tau_g^{\gamma}/2}\right)
  \bigg],
\end{multline}
where $\tau_g^\gamma = \tau-|Z|/V_g^\gamma$, and $j_n(x)$ denotes the
spherical Bessel function of the first kind and the $n$-th order. Therefore,
from Eq.~(\ref{12}), we obtain the EDGF, $\ubG(\vR,\tau)$, in the
following form
\begin{align}\label{30}
  \ubG ={ \Dom \over (2\pi)^3 }\sum_{\gamma = p,s}&\int
  d\vk_{t}\,e^{i(\vk_t\cdot\vR+\kappa^\gamma_0|\hvz\cdot\vR|)}
  \times\ubT_{\gamma,\sgn(\hvz\cdot\vR)}
  \left({\textstyle{\Dom\over 2}\left(\tau-{|\hvz\cdot\vR|\over V_g^\gamma}\right)}\right),
\end{align}
where
\begin{equation}
\ubT_{\gamma,\pm 1}(x) =
    \ubC_{\gamma,\pm 1}j_0(x)-{i\Dom\over 2}\ubD_{\gamma,\pm 1}\,j_1(x).
\end{equation}

In some special cases, the remaining integration over $d\vk_t$ can be
performed analytically. In Sec.~\ref{paraxial}, we consider such a
special case of paraxial propagation. The representation (\ref{30}) is
most suitable for the calculation of fields of sources with a known
spatial spectrum in the transverse plane, e.g.\ in near-field radiative
heat transfer problems. Considering arbitrary source vectors and using
Eqs. (\ref{15}) and (\ref{30}), we can investigate the propagation of
the EMFSVA in any unbounded anisotropic dispersive media.

The present formalism can be extended to contain interface effects,
 which will make it applicable to investigate the
propagation of the EMFSVA via multilayer media. In Sec.~\ref{layered}, we
consider a special case of such media when the neighboring layers are
(approximately) impedance matched.

\section{\label{paraxial}EDGF for paraxial propagation}

Let us consider the case when the envelope propagation happens
dominantly along the anisotropy axis. This case is typical for
extremely anisotropic uniaxial MM in which $|\epz| \gg |\ept|$ and
(or) $|\mz| \gg |\mt|$. Indeed, the propagation factor
$\kappa_0^\gamma$ from Eq.~(\ref{28}) can be expressed as
\begin{equation}
  \kappa^{p[s]}_{0} = \kappa_{a}\sqrt{1-{a^{e[m]}_{t}\kt2 \over a^{e[m]}_{z}\kappa_a^2}},
\end{equation}
where $\kappa_{a}=\sqrt{a^e_{t}a^m_{t}}$. When
$\left|a^{e[m]}_{z}\kt2\right|\ll
\left|a^{e[m]}_{z}\kappa_a^2\right|$, we can expand $\kappa_0^\gamma$
as follows
\begin{equation}
  \label{eq:kappa00}
  \kappa^{p[s]}_{0} \approx \kappa_{a}-{1\over 2}{a^{e[m]}_{t}\kt2\over a^{e[m]}_{z}\kappa_{a}}.
\end{equation}
On the other hand, from Eqs.~(\ref{22}) and~(\ref{23}) it is seen that
in this case the $\ubg_{tt}$ component dominates over the $\ubg_{tz}$,
$\ubg_{zt}$ and $\ubg_{zz}$ components, so that the EDGF is dominantly
transverse which implies that the wave energy propagates dominantly
along the $z$-axis.

Therefore, when calculating the integral over $d\vk_t$ in
Eq.~(\ref{30}), we will not make a big mistake if we evaluate the
$\ubT_{\gamma,\pm 1}$ term of Eq.~(\ref{30}) at $k_t \rightarrow 0$
while using the expansion~(\ref{eq:kappa00}) in the exponential
terms. In this approximation, the integration over $d\vk_t$ can be
performed analytically, which results in the following expression for
the $tt$-block of the paraxial EDGF:
\begin{align}
  \label{paraxEDGF}
  &\ubg_a^{tt}(x-x',y-y',\pm|z-z'|,t-t') =
  { \Dom \over (2\pi)^2 }\sum_{\gamma = p,s}\int\limits_{-\infty}^{\xi_{\gamma}}\!d\xi\,
  \ubt_{\gamma\pm}{e^{i\kappa_{a}\left(|z-z'|+
  {(x-x')^2+(y-y')^2\over 2\xi}\right)}\over 2\xi},
\end{align}
where $\xi_{p[s]} = {a^{e[m]}_t\over a^{e[m]}_z}|z-z'|$, and
\begin{align}
  \ubt_{s\pm} &=  
    \begin{bmatrix}
      -{a_t^m{\hvz\hvz\xx\nabla_t\nabla_t}\over 2\kappa_a} & \pm {\hvz\times\nabla_t\nabla_t\over 2}\\
      \mp {\nabla_t\nabla_t\times\hvz\over 2} & -{a_t^e\nabla_t\nabla_t\over 2\kappa_a}
    \end{bmatrix}
    j_0(\tau)-{i\Dom\over 2}
    \left({b_t^m\over a_t^m}-{b_t^e\over a_t^e}\right)
    \begin{bmatrix}
      -{a_t^m{\hvz\hvz\xx\nabla_t\nabla_t}\over4\kappa_a} & \u0\\
      \u0 & {a_t^e{\nabla_t\nabla_t}\over4\kappa_a}
    \end{bmatrix}
    j_1(\tau),
\end{align}
\begin{align}
  \ubt_{p\pm} &=  
    \begin{bmatrix}
      -{a_t^m{\nabla_t\nabla_t}\over 2\kappa_a} & \pm {\nabla_t\nabla_t\times\hvz\over 2}\\
      \mp {\hvz\times\nabla_t\nabla_t\over 2} & -{a_t^e\hvz\hvz\xx\nabla_t\nabla_t\over 2\kappa_a}
    \end{bmatrix}
    j_0(\tau)-{i\Dom\over 2}
    \left({b_t^m\over a_t^m}-{b_t^e\over a_t^e}\right)
    \begin{bmatrix}
      -{a_t^m{\nabla_t\nabla_t}\over 4\kappa_a} & \u0\\
      \u0 & {a_t^e{\hvz\hvz\xx\nabla_t\nabla_t}\over4\kappa_a}
    \end{bmatrix}
    j_1(\tau),
\end{align}
with $\nabla_t = \hvx(\p/\p x) + \hvy(\p/\p y)$,
$\tau = {\Dom\over 2}\left(t - t' - |z-z'|/V_g^a\right)$, and
$V_g^a = {2\kappa_a/(a_{t}^{e}b^{m}_{t}+a^{m}_{t}b^{e}_{t})}$. The
integral over $d\xi$ in Eq.~(\ref{paraxEDGF}) can be taken in a closed
form after the dyadic differential operators $\ubt_{\gamma\pm}$ have
acted on the exponential term. The integral
representation~(\ref{paraxEDGF}) is especially handy when taking the
convolution integral in Eq.~(\ref{15}) with the source currents having
Gaussian profiles in the $xy$-plane.

\section{\label{layered}Propagation through impedance matched MM layers}

For the sake of this section, we apply the developed formalism for
multilayer uniaxial MMs in which layers are approximately impedance
matched. The anisotropy axis is along the $z$-axis and is the same in
all layers. The interfaces of the layers are perpendicular to the
$z$-axis and are located at planes $z = z_l$, where $l$ is the layer
index. Here, we are interested only in the paraxial propagation.

As the source of excitation, we consider oscillating surface electric
current distributed in the plane $z = z'$ with some amplitude profile
along the $x$-direction. The plane $z=z'$ happens to be inside one of
the layers (which we will call the 0-th layer) located at
$z \in (z_0,z_1)$. In the $y$-direction, the source current is
uniform. The source current density is oscillating in time with the
carrier frequency $\omega_0$ and has the SVA envelope of oscillations,
$\vj^{E}_m$, defined as
\begin{equation}\label{38}
\vj^{E}_m = \sum_{\gamma=p,s}J_{\gamma}\hvu_{\gamma}e^{-{x^2\over 2\sigma_{\gamma}^2}-{t^2\over 2\sigma_t^2}}\delta(z-z^{\prime}), \quad
\end{equation}
where $J_{\gamma}\hvu_{\gamma}$ with $\gamma = p,s$, determines the
initial vectorial amplitudes of the $s$- and $p$-polarized components of the
surface current (here, $\hvu_p=\hvx$ and $\hvu_s=\hvy$),
$\sigma_{p[s]}$ defines the characteristic width of the amplitude
profiles in $x$, separately for the two polarizations, and where we
assume that $\sigma_t$, the envelope duration in time, is such that
${\Dom\over 2\pi}\sigma_t \gtrsim 1$. Under this condition,
practically all source spectral power is concentrated within the
frequency interval of width~$\Dom$. Thus, the SVA of the 6-vector
source current density reads
\begin{equation}
  \begin{aligned}
    \vbJ_m(x,z,t) &= \vbJ_0(x,t)\delta(z-z^{\prime}),\qquad 
    \vbJ_0(x,t) &= \sum_{\gamma=p,s}
  \begin{bmatrix}
    J_{\gamma}\hvu_\gamma\\ 0
  \end{bmatrix}\,e^{-{x^2\over 2\sigma_{\gamma}^2}-{t^2\over
      2\sigma_t^2}}.
  \end{aligned}
  \label{eq:6source}
\end{equation}
Such a source creates the electromagnetic field in the 0-th layer which
propagates in both $z>z'$ and $z<z'$ directions. When this field reaches
the interface $z=z_1$ between the 0th and the next layer, it excites
the fields in the next layer, and so on. If the characteristic wave
impedances of the neighboring layers are mismatched, the reflected field
will also appear, which can propagate to the other interface, be
partially reflected again, etc.

We reserve the study of such multiple reflections in the EDGF context
for a future work. Here, we assume that the neighboring layers are
approximately impedance matched at the frequencies close to $\omega_0$,
and the reflections may be neglected. Note that this does not mean
that the layers must be made of the same materials, or that the
materials must have the same dispersion. The impedance match condition
for the case of the paraxial propagation considered in this section, means
that the material parameters of the $l$-th and $(l+1)$-th layer
satisfy
\begin{equation}
  {\ept\over\mt}\bigg|_{\omega_0,\,l} \approx {\ept\over\mt}\bigg|_{\omega_0,\,l+1}.
\end{equation}

Under this assumption, the EMFSVA created by the
source~(\ref{eq:6source}) in the 0-th layer can be obtained from
Eqs.~(\ref{15}) and (\ref{paraxEDGF}) and expressed in the following
form, after evaluating all involved integrals:
\begin{align}
  &\vbF_m(x,\pm|z-z'|,t;\uep{}^0,\um{}^0,\vbJ_0) = 
  \sum_{\gamma=p,s}{\sigma_{\gamma}\,e^{i\kappa_a|z-z'|-{x^2\over 2\tilde{\sigma}_{\gamma}^2}}\over \tilde{\sigma}_{\gamma}}\,\sum_{n=-\infty}^\infty
  \ubh_{\gamma\pm}(\tau_n)\cdot\vbJ_0(0,t_n),
  \label{eqFJ}
\end{align}
where $\tau_n = {\Delta\omega\over 2}(t - t_n -{|z-z'|\over V_g^a})$,
$\tilde{\sigma}_{p[s]} = \sqrt{\sigma_{p[s]}^2+ {ia^{e[m]}_t|z-z'|\over a^{e[m]}_z\kappa_a}}$, and
\begin{align}
  \ubh_{s\pm}(\tau) &=  
    \begin{bmatrix}
      -{a_t^m\hvy\hvy\over 2\kappa_a} & \pm {\hvy\hvx\over 2}\\
      \pm {\hvx\hvy\over 2} & -{a_t^e\hvx\hvx\over 2\kappa_a}
    \end{bmatrix}
                     j_0(\tau)-{i\Dom\over 2}
                    \left({b_t^m\over a_t^m}-{b_t^e\over a_t^e}\right)
    \begin{bmatrix}
      -{a_t^m\hvy\hvy\over 4\kappa_a} & \u0\\
      \u0 & {a_t^e\hvx\hvx\over 4\kappa_a}
    \end{bmatrix}
    j_1(\tau),
\end{align}
\begin{align}
  \ubh_{p\pm}(\tau) &=  
    \begin{bmatrix}
      -{a_t^m\hvx\hvx\over 2\kappa_a} & \mp {\hvx\hvy\over 2}\\
      \mp {\hvy\hvx\over 2} & -{a_t^e\hvy\hvy\over 2\kappa_a}
    \end{bmatrix}
                     j_0(\tau)-{i\Dom\over 2}
                    \left({b_t^m\over a_t^m}-{b_t^e\over a_t^e}\right)
    \begin{bmatrix}
      -{a_t^m\hvx\hvx\over 4\kappa_a} & \u0\\
      \u0 & {a_t^e\hvy\hvy\over4\kappa_a}
    \end{bmatrix}
    j_1(\tau),
\end{align}
where the parameters $a^{e[m]}_t$, $b^{e[m]}_t$, etc.\ are expressed
through the components of the material dyadics of the 0-th layer,
$\uep{}^0$ and $\um{}^0$.

In order to find the fields in the next layer located at
$z\in (z_1,z_2)$ (the 1st layer), we introduce an equivalent Huygens source (a pair
of electric and magnetic surface currents) placed at $z = z_1$ (where $z_1 > z'$) defined
as 
\begin{align}
  &\vbJ_m(x,z,t) = \vbJ_1(x,t)\delta(z-z_1),\qquad 
  \vbJ_1(x,t) = \begin{bmatrix} \u0 & \hvz\times\uI_t\\ -\hvz\times\uI_t & \u0\end{bmatrix}\cdot
  \vbF_m(x,z_1-z',t;\uep{}^0,\um{}^0,\vbJ_0).\label{eqJF}
\end{align}
The field in the layer $z \in (z_1,z_2)$ can be found from
Eq.~(\ref{eqFJ}) as $\vbF_m(x,z-z_1,t;\uep{}^1,\um{}^1,\vbJ_1)$, where
$\uep{}^1$ and $\um{}^1$ are the material dyadics of the 1st layer,
from which the equivalent Huygens source $\vbJ_2$ at $z = z_2$ is
expressed by Eq.~(\ref{eqJF}) through
$\vbF_m(x,z_2-z_1,t;\uep{}^1,\um{}^1,\vbJ_1)$, etc. This procedure is
repeated as many times as there are layers at $z>z'$, after which the
fields in the layers located at $z < z'$ can be found in a completely
analogous way.

\section{\label{sec:exampl}Numerical examples}

\subsection{Negative refraction and focusing by uniaxial MM with hyperbolic dispersion}

\begin{figure*}[tb]
    \centering
    \includegraphics[width=0.45\linewidth]{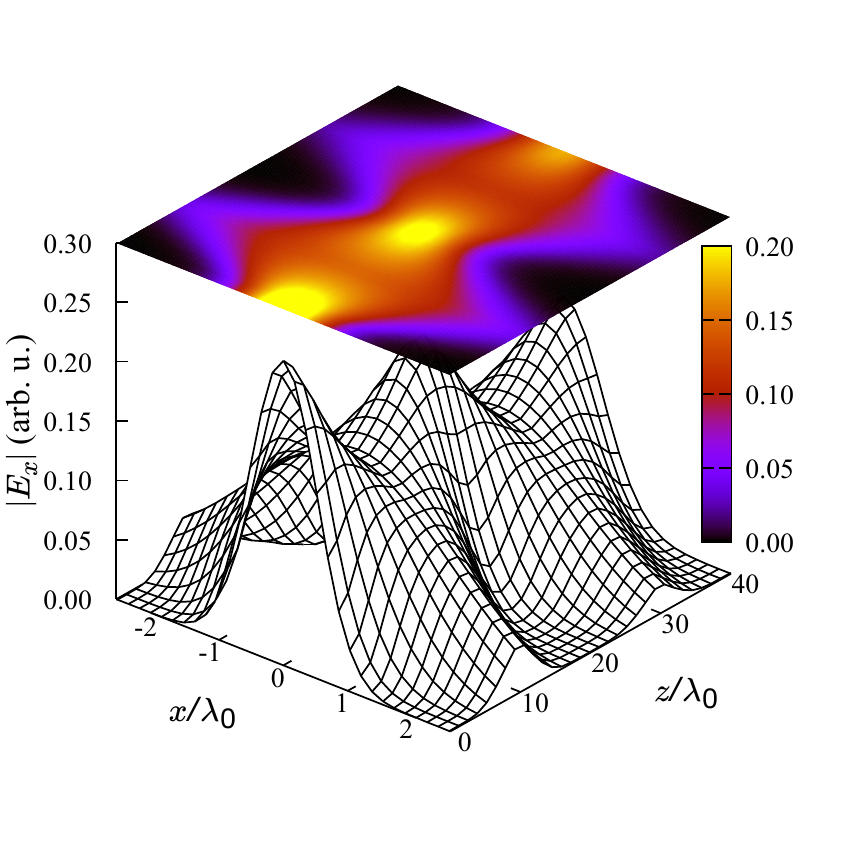}\includegraphics[width=0.45\linewidth]{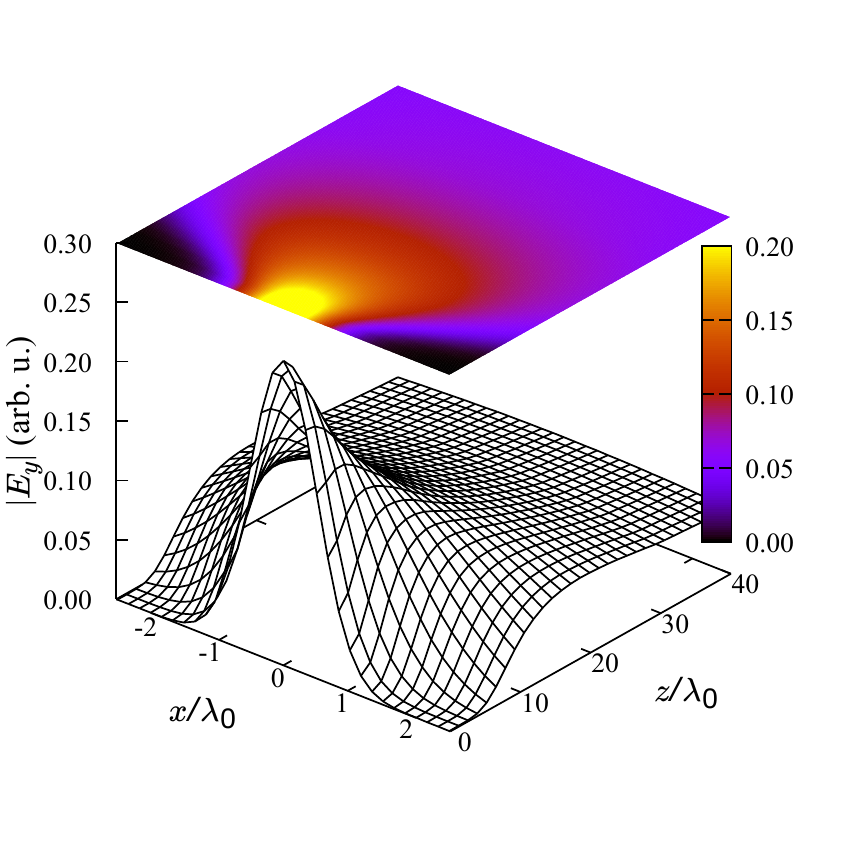}
    \caption{\label{hyperb} (Color online) Paraxial propagation of the
      EMFSVA produced by a source with the Gaussian amplitude profile
      [Eq.~(\ref{38})] through a layer of hyperbolic MM sandwiched
      between two isotropic dielectric layers. Left: Amplitude profile
      of the $p$-polarized beam. Right: Same for the $s$-polarized
      beam.}
\end{figure*}

When the transverse and longitudinal components of the permittivity
dyadic of a MM have opposite signs, the isofrequency curves for the
$p$-polarized waves are hyperbolas. It is known that the $p$-polarized
light refracts negatively when impinging at the interface of a
conventional material and such a hyperbolic MM \cite{Lindell01,Smith04}. In the following
numerical example, we study the implications of this phenomenon on
the paraxial propagation of the Gaussian envelopes considered in
Sec.~\ref{layered}. We shall confirm that the EDGF formalism correctly
predicts focusing of a diverging $p$-polarized Gaussian beam by the
hyperbolic MM.

At near-infrared frequencies, a hyperbolic MM can be realized, e.g.\
by embedding vertically aligned metallic nanowires into an isotropic
dielectric host. For the sake of a numerical example, here we consider
a MM formed by golden nanowires embedded into alumina substrate. By
using the Maxwell-Garnett effective medium theory (EMT) for a uniaxial
MM formed by such nanowires, the following expressions for the
effective transverse and axial permittivities can be obtained~\cite{Starko15}:
\begin{align}
  \label{epsefft}
  \vep_{{\rm eff},t} &= \vep_{h}{\vep_m(1+f) + \vep_h(1-f)\over \vep_m(1-f) + \vep_h(1+f)},\\
  \label{epseffz}
  \vep_{{\rm eff},z} &= \vep_mf + \vep_h(1 - f),
\end{align}
where $\vep_h$ and $\vep_m$ are the dielectric permittivities of the host
material (Al$_2$O$_3$~\cite{Kishkat12}) and the plasmonic metal (Au),
respectively, and $f$ is the nanowires volume fraction. The relative
permittivity of gold at near-infrared frequencies follows the Drude
dispersion model~\cite{Olmon12}
\begin{equation}
  \vep_{m}(\omega) = 1 - {\omega_{p}^2\over
    \omega(\omega+i\tau_D^{-1})},
\end{equation}
where $\hbar\omega_{p} = 8.5$~eV and $\tau_D = 1.4\times 10^{-14}$~s.
At the frequencies below the plasma frequency $\omega_p$,
$\Re(\vep_m(\omega)) < 0$, and one can achieve
$\Re(\vep_{{\rm eff},z}(\omega)) < 0$ with a proper choice of the
nanowires volume fraction $f$.

Let us consider a structure comprised of a hyperbolic MM sandwiched
between two isotropic dielectrics with relative permittivity
$\vep_d = 3.8$ (e.g.\ aluminum nitride~\cite{Kishkat12}). The volume
fraction of Au nanowires embedded into Al$_2$O$_3$ substrate is
$f = 0.15$.  The carrier frequency is set to
$\hbar\omega_0=1.32$~eV. An oscillating electric current source with
this frequency and the amplitude profile given by Eq.~(\ref{38}) with
$\sigma_{p,s}=\lambda_0/\sqrt{\vep_d}$ (where
$\lambda_0=2\pi c/\omega_0$) and $\sigma_t\rightarrow\infty$ is placed
inside the first dielectric at $z = 0$. The hyperbolic MM layer is
located at $10 < z/\lambda_0 < 30$. At the frequency $\omega_0$, the
relative transverse permittivity of this layer is
$\vep_{{\rm eff},t}=3.8+i6.5\times10^{-3}$ and the relative axial
permittivity is $\vep_{{\rm eff},z}=-3.8+i0.22$ [Eqs.~(\ref{epsefft})
and (\ref{epseffz})]. The results of the paraxial EDGF-based
calculations for this structure for the $p$- and $s$-polarized source
currents are displayed in Fig.~\ref{hyperb}. As one can see, the
initially diverging $p$-polarized Gaussian beam, after refracting
negatively at the interface $z/\lambda_0 = 10$, is focused at
$z/\lambda_0 = 20$, the middle point of the hyperbolic MM layer, and
then it diverges again. After reaching the second interface at
$z/\lambda_0=30$, the beam undergoes another negative refraction and
is focused inside the second dielectric layer at the point
$z/\lambda_0 = 40$. On the contrary, the $s$-polarized beam does not
experience any refraction at the MM interfaces and simply
diverges. Note that the scales on the $x$-axis and the $z$-axis in
Fig.~\ref{hyperb} are different, so that the field profile along $z$
is compressed in comparison with that along $x$.

In order to explain how our EDGF formalism is able to reproduce these
phenomena in the paraxial approximation, let us consider the
expression for the square of the effective beam width,
$\tilde{\sigma}_{p[s]}^2$, at a point $z = z''$ inside the MM layer
\begin{align}
  \tilde{\sigma}_{p[s]}^2\big|_{z = z''} &= \tilde{\sigma}_{p[s]}^2\big|_{z = z_1} + i{a^{e[m]}_t|z''-z_1|\over
  a^{e[m]}_z\kappa_a} 
  = \sigma_{p[s]}^2 + i\left({|z_1|\over\kappa_{a,d}}+{a^{e[m]}_t|z''-z_1|\over
      a^{e[m]}_z\kappa_a}\right),
\end{align}
as follows from Eq.~(\ref{eqFJ}). Here, $z_1 = 10\lambda_0$ is the
coordinate of the first MM interface and
$\kappa_{a,d} = 2\pi\sqrt{\vep_d}/\lambda_0\approx\kappa_a$ is the propagation factor
in the dielectric layer. In the considered hyperbolic MM,
$a^{e}_t/a^{e}_z \approx -1$, while $a^{m}_t/a^{m}_z = +1$. Therefore,
when $|z''-z_1| = |z_1|$, for the $p$-polarization, the propagation in the dielectric is
compensated by the propagation in the MM and thus
$\tilde{\sigma}_{p}^2\big|_{z = z''} \approx \sigma_{p}^2$, i.e., the
$p$-polarized Gaussian beam is refocused at the middle of the MM
layer.

\subsection{\label{sec:superlum}Negative group velocity and superluminality in active media}

In this example, we apply our EDGF formalism to the EMFSVA propagation
through a layered structure which is formed by an active medium
sandwiched between two passive media. The active layer is the
$^{132}$Xe gas with inverted population, and the passive media are
air. The EMFSVA is created by an $s$-polarized surface electric
current source located at $z = 0$ with the amplitude profile defined
by Eq.~(\ref{38}) in which $\sigma_s \rightarrow \infty$ (i.e., only
$k_t=0$ component is present). This source creates the EMFSVA
propagating through the three media in the $z>0$ direction. Because
the relative permittivities and permeabilities of the layers are
rather close to unity, the layers are well impedance matched and we
can apply the theory of Sec.~\ref{layered}.

The relative permittivity of the $^{132}$Xe gas with inverted population
follows the Lorentzian dispersion with negative oscillator
strength~\cite{47}:
\begin{equation}\label{25}
\varepsilon^{L}_t = \varepsilon^{L}_z = 1 - {\eta\omega_{p}^{2} \over \omega_{r}^{2} - \omega_{0}^{2} - i\gamma\omega_{0} },
\end{equation}
where the parameter $\omega_{p}/2\pi = 0.42$~GHz accounts for both the
magnitude of the oscillator strength and the atomic plasma frequency,
$\eta = 0.9$ is the relative inversion, and $\omega_{r}/2\pi = 84$~THz
and $\gamma/2\pi = 4.2$~MHz are the resonant frequency and the
linewidth, respectively. We stay detuned from the resonant frequency
and set $\omega_{0} = \omega_{r} + \omega_{p}/3$. At this point, the
group velocity in the active layer is $V_g^s \approx -0.97c$ and the
assumptions of the EDGF approach hold (see Appendix~C for details).

\begin{figure*}[tb]
	\centering
	\includegraphics[width=0.9\linewidth]{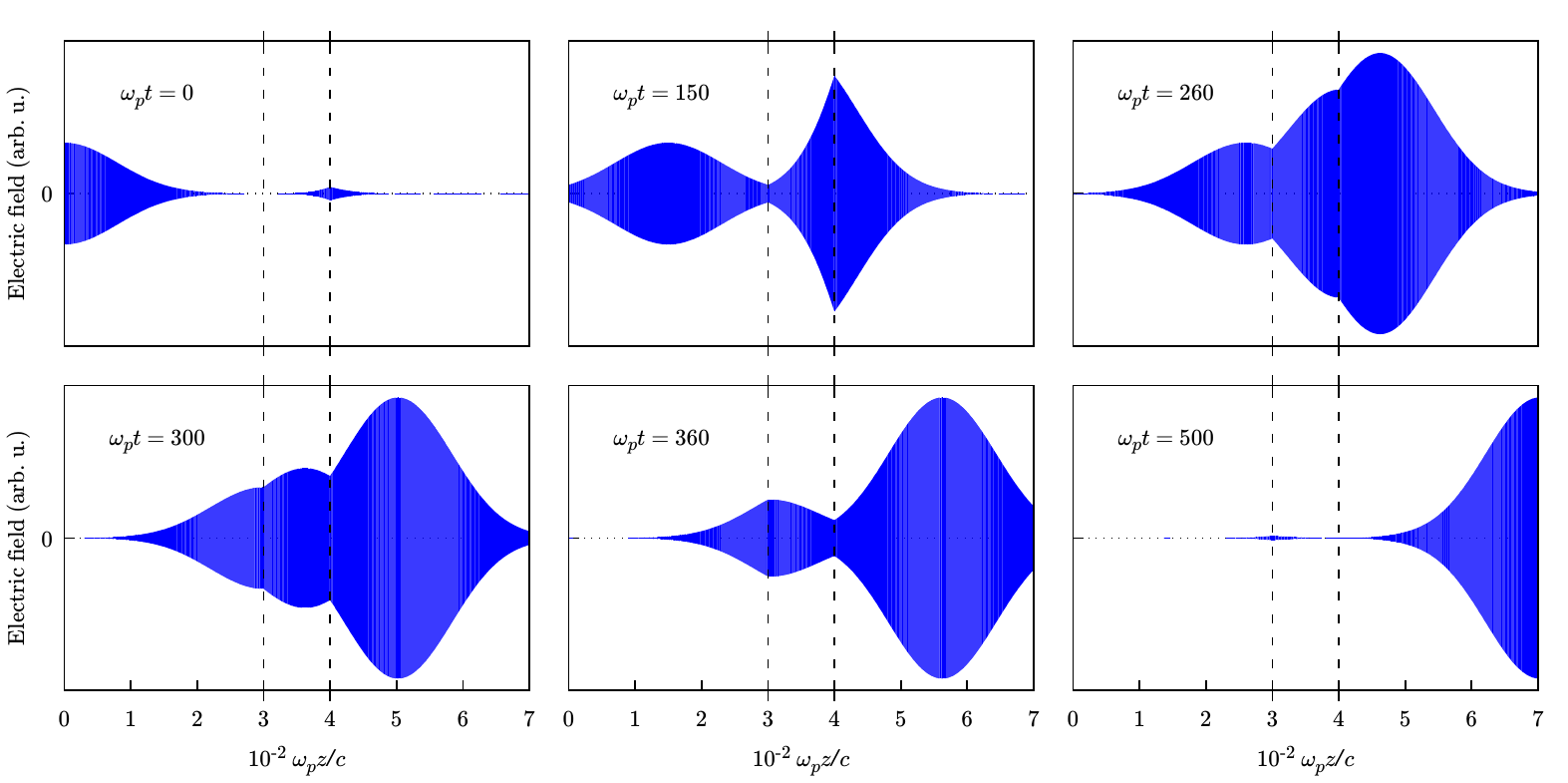} 
	\caption{ (Color online) Superluminal propagation of the
          EMFSVA envelope with normalized duration
          $\omega_p\sigma_t = 80$ through a three-layer medium
          comprised of two air layers at $\omega_pz/c<300$ and
          $\omega_pz/c>400$, and an inverted population $^{132}$Xe gas
          layer placed in between, plotted at various normalized times
          $\omega_pt$ versus normalized axial coordinate
          $\omega_pz/c$ (for further details, see
          Sec.~\ref{sec:superlum}).} \label{R2}
\end{figure*}

In Fig.~\ref{R2}, we depict the propagation of the EMFSVA envelope
with $\omega_p\sigma_t = 80$ through the three layers (separated by
dashed vertical lines in the figure) at various moments of the
normalized time: $\omega_pt =$ 0, 150, 260, 300, 360, 500. The
observed behavior agrees with that of Ref.~\cite{47}. As is seen from
the top plots in Fig.~\ref{R2}, at $\omega_pt = 0$, the front edge
(the precursor) of the Gaussian envelope penetrates into the second
layer (the active layer). At $\omega_pt= 150$, the maximum of the
original pulse passes forward and the field acquires a noticeable
value at the interface of the first and the second layers and, at the
same time, we can see that the amplified field in the active layer
forms a sharp peak and penetrates into the third layer. We can see how
the back propagating pulse is formed in the gain medium and how it
interacts with the primary pulse which propagates forward in the first
layer at the same interface at $\omega_pt = 260, 300,$ and
$360$. Finally, at $\omega_pt = 500$, we can see that the main pulse
has left the two layers, however, a small effect of its back edge is
still present at the interface of the first and the second layers. It
should be noted that at $\omega_pt = 260$ the maximum of the pulse
exits the second layer earlier than it would do if it had traveled
through an equal distance of air, i.e.~it appears superluminal. This
happens due to the action of the gain medium on the electromagnetic
field with a Gaussian-shaped envelope which lacks a definite turn-on
moment. Analogous results are reported in Ref.~\cite{47}.

\section{\label{sec:concl}Conclusions}

In this work, we have presented a theoretical formalism which is
applicable for studying the propagation of amplitude fluctuations of
the quasi-monochromatic electromagnetic field through anisotropic
dispersive media. The developed formalism is aimed to be used in
future works to model the dynamics of RHT in such media, in
particular, in uniaxial MMs~\cite{22}, however, it is equally
applicable to the analysis of narrow-band signal propagation in these
MMs.

Starting with the 6-vector Maxwell's operator equation, we have
formulated an EDGF-based method with which we have derived the
envelope Green's functions used to calculate the EMFSVA propagating
through a dispersive medium with uniaxial dyadic constitutive
parameters. We have obtained the matrix elements of the EDGF in the
Fourier and the configuration spaces for the considered media. In the
case of paraxial propagation, the EDGF for uniaxial media can be
written in a closed form, resulting in a formulation analogous to the
Gaussian beam-based paraxial approximation in optics. Finally, we have
considered propagation of the EMFSVA through non-magnetic passive and
active layered media. We simulated the propagation of the EMFSVA
through such layered media by employing the effective Huygens sources
at the interfaces of the neighboring layers.

We have demonstrated with numerical examples that the developed
formalism correctly models negative refraction and focusing by
hyperbolic MMs and is also applicable to exotic effects in optical
media with inverted population, such as the negative group velocity
and superluminality. The group velocity obtained in our formalism
agrees with the standard definition known from the literature and
results in similar behaviors. The considered examples confirm the
applicability and validity of the EDGF approach developed in this
paper.

\section*{Acknowledgment}
The authors acknowledge support under the project
Ref.~UID/EEA/50008/2013, sub-project SPT, financed by Funda\c{c}\~{a}o
para a Ci\^{e}ncia e a Tecnologia (FCT)/Minist\'{e}rio da Ci\^{e}ncia,
Tecnologia e Ensino Superior (MCTES), Portugal. S.I.M. acknowledges
support from Funda\c{c}\~{a}o para a Ci\^{e}ncia e a Tecnologia (FCT), Portugal,
under Investigador FCT~(2012) grant
(Ref.\ IF/01740/2012/CP0166/CT0002).

\appendix

\section*{Appendix A}

In its standard formulation, Kotelnikov's theorem expresses the signal
$s(t)$ with a limited spectrum,
$\Omega \in [-{\Delta\omega\over2},{\Delta\omega\over2}]$, through a set of the
discrete samples, $s(t_n)$:
\begin{equation}
s(t) = \sum_{n = -\infty}^{+\infty}s(t_n)j_0\left({\Delta\omega\over2}(t-t_n)\right), \quad t_n =
{2\pi n\over \Delta\omega},
\end{equation}
where $j_0(x) = \sin(x)/x$. Using this theorem, the convolution
integral of two functions, $a(t)$ and $s(t)$, both satisfying
Kotelnikov's spectral condition, can be written as follows
\begin{equation}
  u(t) = \int\limits_{-\infty}^{+\infty}a(t-t')s(t')\,dt' = {2\pi\over\Delta\omega}\sum_{n=-\infty}^{+\infty}a(t-t_n)s(t_n).
\end{equation}
We have used this result when obtaining Eq.~(\ref{15}).

\section*{Appendix B}

Regarding Eq.~(\ref{defCandD}), the different blocks of
$\ubC_{\gamma,\pm1}$ and $\ubD_{\gamma,\pm1}$ ($\gamma = p, s$)
matrices are given as follows
\begin{equation}
\ubC_{\gamma,{\rm sgn}(Z)}=
\begin{bmatrix}
\uC^{tt,\gamma} && \uC^{tz,\gamma} \\
\uC^{zt,\gamma} && \uC^{zz,\gamma}
\end{bmatrix}, \quad
\ubD_{\gamma,{\rm sgn}(Z)}=
\begin{bmatrix}
\uD^{tt,\gamma} && \uD^{tz,\gamma} \\
\uD^{zt,\gamma} && \uD^{zz,\gamma}
\end{bmatrix}.
\end{equation}
Using Eqs.~(\ref{22}) and (\ref{23}), recalling $\kappa^{\gamma}_{0}$
and $V_{g}^{\gamma}$ from Eqs.~(\ref{28}) and (\ref{29}),
respectively, we obtain the non-zero
components of $\uC$'s and
$\uD$'s as follows [the structure of these
sub-blocks is the same as defined by Eqs.~(\ref{gtt})--(\ref{gzz})]:
\begin{gather}
C^{tt,p[s]}_{ee[mm]}=-{a^{m[e]}_{t} - k_{t}^2/a^{e[m]}_{z} \over 2\kappa^{p[s]}_{0}}, \qquad
C^{tt,p[s]}_{mm[ee]}=-{a^{e[m]}_{t} \over 2\kappa^{p[s]}_{0}}, \qquad C^{tt,s[p]}_{em} =[-]{\mbox{sgn}(Z)\over 2},\\
C^{tz,p[s]}_{ee[mm]}={\mbox{sgn}(Z)k_{t}\over 2a^{e[m]}_{z}}, \qquad
C^{tz,p[s]}_{me[em]}=[-]{a^{e[m]}_{t}k_{t} \over 2 a^{e[m]}_{z}\kappa^{p[s]}_{0}},\\
C^{zz,p[s]}_{ee[mm]}=-{a^{e[m]}_{t} k^{2}_{t} \over 2{a_{z}^{e[m]}}^2 \kappa^{p[s]}_{0}},
\end{gather}
\begin{gather} 
  D^{tt,p[s]}_{ee[mm]}=-{b_t^{m[e]}+k_t^2b_z^{e[m]}/{a_z^{e[m]}}^2\over
    2\kappa_0^{p[s]}} +
  {V_g^{p[s]}}^{-1}{a_t^{m[e]}-k_t^2/a_z^{e[m]}\over
    2{\kappa_0^{p[s]}}^2},\qquad
  D^{tt,p[s]}_{mm[ee]}=-{b_t^{e[m]}\over
    2\kappa_0^{p[s]}}+{V_g^{p[s]}}^{-1}{a_t^{e[m]}\over
    2{\kappa_0^{p[s]}}^2}, \qquad D^{tt,s[p]}_{em}=0,\\
  D^{tz,p[s]}_{ee[mm]}=-{\mbox{sgn}(Z)b^{e[m]}_{z} k_{t}\over {2 a_{z}^{e[m]}}^2},\qquad
  D^{tz,p[s]}_{me[em]}=[-]{k_ta_t^{e[m]}\over
	2\kappa_0^{p[s]}a_z^{e[m]}}\left({b_t^{e[m]}\over
	a_t^{e[m]}}-{b_z^{e[m]}\over a_z^{e[m]}}-{{V_g^{p[s]}}^{-1}\over
	\kappa_0^{p[s]}}\right), \\
    D^{zz,p[s]}_{ee[mm]}={k_t^2a_t^{e[m]}\over
  	2\kappa_0^{p[s]}{a_z^{e[m]}}^2}\left(2{b_z^{e[m]}\over
  	a_z^{e[m]}}-{b_t^{e[m]}\over a_t^{e[m]}}+{{V_g^{p[s]}}^{-1}\over \kappa_0^{p[s]}}\right)\\
  \uC^{zt,\gamma} = \left(\uC^{tz,\gamma}\right)^{T}, \qquad \uD^{zt,\gamma} = \left(\uD^{tz,\gamma}\right)^{T}.
\end{gather}

\section*{Appendix C}

In order to check if it is reasonable to ignore the $O(\Omega^{2})$
term in Eq.~(\ref{26}), we consider Eq.~(\ref{20}) with an extra
second-order term:
\begin{equation}
  d^{e[m]}_l = a^{e[m]}_l+b^{e[m]}_l\Omega + {c^{e[m]}_l\Omega^2\over 2},
\end{equation}
where $l = t$ or $z$, and substitute it into Eq.~(\ref{24}). We define
the smallness parameter $\delta_{p[s]}$ as the ratio of the second-order $O(\Omega^2)$
term [which is dropped in Eq.~(\ref{26})] to the first-order
$O(\Omega)$ term in the Taylor expansion of Eq.~(\ref{24}) as follows
\begin{equation}
  \delta_{p[s]} = {\Omega V_g^{p[s]}\over 2}\left.{\partial^2\kappa_{p[s]}\over\partial\Omega^2}\right|_{\Omega = 0},
\end{equation}
where in the following calculations we replace $\Omega$ with its
maximum value $\Omega = \Dom/2$. Obviously, $\delta_{p[s]}$ depends on
$\omega_{0}$ and $\Dom$, in addition to the material
properties. Considering $k_{t}=0$, when both polarizations are
equivalent, we obtain
\begin{equation}
  \delta_a = {\Delta\omega\over 4\kappa_a}\left[{V_g^a}\left({a_t^ec_t^m+a_t^mc_t^e\over 2}+b_t^eb_t^m\right)-{1\over
  V_g^a}\right].
\end{equation}

\begin{figure}[htb]
	\begin{center}
		\includegraphics[width=0.6\linewidth]{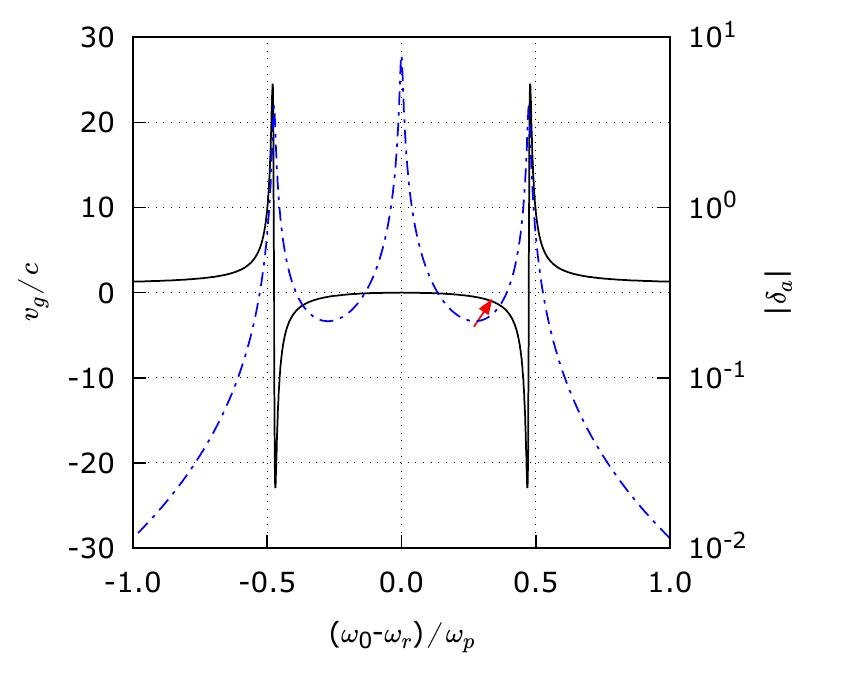}
		\caption{(Color online) The normalized group velocity
			$\upsilon_g/c$ (black solid curve) and the smallness
			parameter magnitude $|\delta_a|$ (blue dashed curve)
			as functions of the normalized frequency shift
			$(\omega_0 - \omega_r)/\omega_p$. The red arrow
			indicates the operating point corresponding to the
			pulse propagation shown in Fig.~\ref{R2}. The
			material parameters are the same as in
			Sec.~\ref{sec:superlum}.}
		\label{SVg}
	\end{center}
\end{figure}

In Fig.~\ref{SVg}, we depict the normalized group velocity and the
smallness parameter versus the normalized frequency shift (for the
example of Sec.~\ref{sec:superlum}). In this figure,
$\upsilon_{g}=\Re(V_{g}^a)$ and $\delta_{a}$ are measured by the
scales on the left and the right, respectively. As can be seen from
Fig.~\ref{SVg}, $|\delta_{a}|\approx 0.2$ near the operational
frequency $\omega_0 = \omega_r + \omega_p/3$ from the example of
Sec.~\ref{sec:superlum}. This value can be considered sufficiently
small. On the other hand, in the regions where the group velocity
becomes extremely superluminal, we can see that $|\delta_{a}| \gg 1$,
which indicates that in these regions the pulse propagation is very
much affected by the group dispersion and the group velocity looses
its physical meaning.

\pagebreak


\begin{thebibliography}{99}
\bibitem{1} H. Lamb, {\it Hydrodynamics} (Cambridge, University Press, 1916).
\bibitem{2} A. Schuster, {\it An introduction to the theory of optics} (London, Edward Arnold, 1904). 
\bibitem{3} L.I. Mandelshtam, {\it Zh. Eksp. Teor. Fiz.} {\bf 15}, 475 (1945). 
\bibitem{4} V.G. Veselago, {\it Sov. Phys. Uspekhi.} {\bf 10}, 509 (1968).
\bibitem{5} S. Wuestner, A. Pusch, K.L. Tsakmakidis, J.M. Hamm, and
  O. Hess, {\it Phys. Rev. Lett.} {\bf 105}, 127401 (2010).
\bibitem{5a} D. Ye, K. Chang, L. Ran, and H. Xin, {\it Nat. Comm.}
  {\bf 5}, 5841 (2014).
\bibitem{6} D.R. Smith, J.B. Pendry, and M.C.K. Wiltshire, {\it Science} {\bf 305}, 788 (2004).
\bibitem{11} Z. Liu, H. Lee, Y. Xiong, C. Sun, and X. Zhang, {\it Science} {\bf 315}, 1686 (2007).
\bibitem{8} J.B. Pendry, D. Schurig, and D.R. Smith, {\it Science} {\bf 312}, 1780 (2006).
\bibitem{19} T.A. Morgado, J.S. Marcos, S.I. Maslovski, and
  M.G. Silveirinha, {\it Appl. Phys. Lett.} {\bf 101}, 021104
  (2012).
\bibitem{12} D. Schurig, J.J. Mock, B.J. Justice, S.A. Cummer,
  J.B. Pendry, A.F. Starr, and D.R. Smith, {\it Science} {\bf 314}, 977
  (2001).
\bibitem{9} N.I. Landy, S. Sajuyigbe, J.J. Mock, D.R. Smith, and
  W.J. Padilla, {\it Phys. Rev. Lett.} {\bf 100}, 207402 (2008).
\bibitem{10} C.A. Valagiannopoulos, J. Vehmas, C.R. Simovski,
  S.A. Tretyakov, and S.I. Maslovski, {\it Phys. Rev. B} {\bf 92},
  245402 (2015).
\bibitem{7} U. Leonhardt, Nature {\bf 415}, 406 (2002).
\bibitem{13} H.-T. Chen, W.J. Padilla, J.M.O. Zide, A.C. Gossard,
  A.J. Taylor, and R.D. Averitt, {\it Nature} {\bf 444}, 597 (2006).
\bibitem{14}T. Driscoll, H.-T. Kim, B.-G. Chae, B.-J. Kim, Y.-W. Lee,
  N.M. Jokerst, S. Palit, D.R. Smith, M. Di Ventra, and D.N. Basov,
  {\it Science} {\bf 325}, 1518 (2009). 
\bibitem{15} N.-H. Shen, M. Massaouti, M. Gokkavas, J.-M. Manceau,
  E. Ozbay, M. Kafesaki, T. Koschny, S. Tzortzakis, and
  C.M. Soukoulis, {\it Phys. Rev. Lett.} {\bf 106}, 037403 (2011).
\bibitem{16} S.I. Maslovski and M.G. Silveirinha, {\it Phys. Rev. A}
  {\bf 83}, 022508 (2011). 
\bibitem{18} I.V. Shadrivov, P.V. Kapitanova, S.I. Maslovski, and
  Y.S. Kivshar, {\it Phys. Rev. Lett.} {\bf 109}, 083902 (2012).
\bibitem{20} I. Latella, S.-A. Biehs, R. Messina, A.W. Rodriguez, and
  P. Ben-Abdallah, {\it Phys. Rev. B} {\bf 97}, 035423 (2018). 
\bibitem{21} S.I. Maslovski, C.R. Simovski, S.A. Tretyakov, {\it New
    J. Phys.} {\bf 18}, 013034 (2016).
\bibitem{22} H. Mariji and S.I. Maslovski, in {\it Proceedings of SPIE
    Photonics Europe, 10671, Metamaterials XI}, edited by
  A.D. Boardman, A.V. Zayats, and K.F. MacDonald (SPIE,
  Strasbourg, 2018), p.~1067114. 
\bibitem{23} C. Simovski, S. Maslovski, I. Nefedov, and S. Tretyakov,
  {\it Opt. Express} {\bf 21}(12), 14988 (2013).

\bibitem{24} M.S. Mirmoosa, S.-A. Biehs, and C.R. Simovski, {\it
    Phys. Rev. Applied} {\bf 8}, 054020 (2017).


\bibitem{Wang2000} L.J. Wang, A. Kuzmich, and A. Dogariu, {\it Nature} {\bf 406}, 277 (2000).
\bibitem{26} K.L. Tsakmakidis, T.W. Pickering, J.M. Hamm, A.F. Page,
  and O. Hess, {\it Phys. Rev. Lett.}  {\bf 112}, 167401 (2014).
\bibitem{47} E.L. Bolda, J.C. Garrison, and R.Y. Chiao, {\it Phys. Rev. A}
  {\bf 49}(4), 2938 (1994). 
\bibitem{27} M.S. Bigelow, N.N. Lepeshkin, R.W. Boyd, {\it Science} {\bf 301}, 200 (2003).
\bibitem{28} A.D. Neira, G.A. Wurtz, and A.V. Zayats, {\it Nature} {\bf 5}, 17678 (2015).
\bibitem{29} L. Brillouin, {\it Wave Propagation and Group Velocity} (Academic Press, New York, 1960).
\bibitem{30} A. Kuzmich, A. Dogariu, L. J. Wang, P. W. Milonni, and R. Y. Chiao
{\it Phys. Rev. Lett.} {\bf 86}, 3925 (2001).
\bibitem{31} M.I. Stockman, Phys. Rev. Lett. {\bf 98}, 177404 (2007). 
\bibitem{32} D. Forcella, C. Prada, R. Carminati, {\it Phys. Rev. Lett.} {\bf 118}, 134301 (2017).
\bibitem{33} M.M. Kash, V.A. Sautenkov, A.S. Zibrov, L. Hollberg,
  G.R. Welch, M.D. Lukin, Y. Rostovtsev, E.S. Fry, and
  M.O. Scully, {\it Phys. Rev. Lett.} {\bf 82}, 5229 (1999).
\bibitem{34} L.M. Duan, M.D. Lukin, J.I. Cirac, and P. Zoller, {\it
    Nature} {\bf 414}, 413 (2001).
\bibitem{35} M.D. Lukin and A. Imamo\v{g}lu, {\it Nature} {\bf 413}, 273 (2001).
\bibitem{36} C. Liu, Z. Dutton, C.H. Behroozi, and L.V. Hau, {\it Nature} {\bf 409}, 490 (2001). 
\bibitem{37} D.F. Phillips, A. Fleischhauer, A. Mair,
  R.L. Walsworth, and M.D. Lukin, {\it Phys. Rev. Lett.} {\bf 86}, 783 (2001).
\bibitem{38} M.S. Shahriar, G.S. Pati, R. Tripathi, V. Gopal,
  M. Messall, and K. Salit, {\it Phys. Rev. A} {\bf 75}, 053807 (2007).
\bibitem{39} M. Bajcsy, S. Hofferberth, V. Balic, T. Peyronel,
  M. Hafezi, A.S. Zibrov, V. Vuletic, and M.D. Lukin, {\it Phys. Rev. Lett.} {\bf 102}, 203902 (2009).
\bibitem{40} M. Lee, M.E. Gehm, and M.A. Neifeld, {\it J. Opt.} {\bf 12}, 10 (2010).
\bibitem{41} S. Hrabar, I. Krois, I. Bonic, and A. Kiricenko, {\it Appl. Phys. Lett.} {\bf 102}, 054108 (2013).
\bibitem{42} M. Khorasaninejad, W.T. Chen, J. Oh, and F. Capasso, {\it
    Nano Lett.} {\bf 16}, 3732 (2016).
\bibitem{43} H.N.S. Krishnamoorthy, Z. Jacob, E. Narimanov,
  I. Kretzschmar, and V.M. Menon, {\it Science} {\bf 336}, 205 (2012).
\bibitem{44} G.A. Wurtz, R. Pollard, W. Hendren, G.P. Wiederrecht,
  D.J. Gosztola, V.A. Podolskiy, and A.V. Zayats {\it Nat. Nanotech.} {\bf 6}, 107
  (2011).
\bibitem{45} A. Poddubny, I. Iorsh, P. Belov, and Y. Kivshar, {\it Nat. Phot.} {\bf 7}, 948 (2013).
\bibitem{Lee83} J.K. Lee and J.A. Kong, {\it Electromagnetics} {\bf 3}(2), 111
  (1983).
\bibitem{Lakhtakia89} A. Lakhtakia, V.V. Varadan, and V.K. Varadan,
  {\it Appl. Opt.} {\bf 28}(6), 1049 (1989).
\bibitem{Weiglhofer90} W.S. Weiglhofer, {\it IEE Proc.} {\bf 137}(1), 5 (1990).
\bibitem{Weiglhofer93} W.S. Weiglhofer, {\it Radio Sci.} {\bf 28}(5), 847
  (1993).
\bibitem{Weiglhofer94} W.S. Weiglhofer, {\it Internat. J.
  Electronics}, {\bf 77}(1), 105 (1994).
\bibitem{Lindell99} I.V. Lindell and F. Olyslager,
  {\it J. Electromag. Waves and Appl.} {\bf 13}, 429 (1999).
\bibitem{Olyslager01} F. Olyslager, {\it IEEE Trans. Antennas
  Propagat.} {\bf 49}(4), 660 (2001).
\bibitem{Olyslager02} F. Olyslager and I.V. Lindell, {\it IEEE Antennas
  Propagat. Magazine} {\bf 44}(2), 48 (2002).
\bibitem{Lindell01} I.V. Lindell, S.A.  Tretyakov, K.I.  Nikoskinen,
  and S. Ilvonen, {\it Microwave Opt. Technol. Lett.} {\bf 31}, 129 (2001).
\bibitem{Smith04} D.R. Smith, P. Kolinko, and D. Shurig,
  J. Opt. Soc. Am. B {\bf 21}(5) (2004).
\bibitem{Liu2013} X.L. Liu, R.Z. Zhang, and Z.M. Zhang, {\it
    Appl. Phys. Lett.} {\bf 103}, 213102 (2013).
\bibitem{Kotelnikov33} V.A. Kotelnikov, "On the Capacity of the
'Ether' and Cables in Electrical Communication," {\it Proc. 1st
 All-Union Conf. Technological Reconstruction of the Commun. Sector
 and Low-Current Eng.}, (U.S.S.R., Moscow, 1933).
\bibitem{Starko15} R. Starko-Bowes, J. Atkinson, W. Newman, H. Hu, T. Kallos,
    G. Palikaras, R. Fedosejevs, S. Pramanik, and Z. Jacob,
    {\it J. Opt. Soc. Am. B} {\bf 32}(10), 2074 (2015).
\bibitem{Kishkat12} J. Kischkat, S. Peters, B. Gruska, M. Semtsiv,
    M. Chashnikova, M. Klinkm\:{u}ller, O. Fedosenko, S. Machulik,
    A. Aleksandrova, G. Monastyrskyi, Y. Flores, and W. T. Masselink,
    {\it Appl. Opt.} {\bf 51}(28), 6789 (2012).
\bibitem{Olmon12} R.L. Olmon, B. Slovick, T.W. Johnson, D. Shelton,
    S.-H. Oh, G.D. Boreman, and M.B. Raschke, {\it Phys. Rev. B} {\bf 86},
    235147 (2012).
\end{thebibliography}
\end{document}